\documentclass[aps,pra,superscriptaddress,twocolumn,color,]{revtex4-1}
\usepackage[colorlinks=true, letterpaper=true, pdfstartview=FitV, linkcolor=blue, citecolor=blue, urlcolor=blue]{hyperref}

\usepackage{amsmath}
\usepackage{mathtools, nccmath}
\usepackage{amssymb}
\usepackage{amsfonts}
\usepackage{bm}
\usepackage{graphicx}
\usepackage{array,color}
\usepackage{braket}
\usepackage[utf8]{inputenc}

\begin{document}
\title{Faraday patterns in spin-orbit coupled Bose-Einstein condensates}
\author{Huan Zhang}
\affiliation{
CAS Key Laboratory of Quantum Information,University of Science and Technology of China, Hefei, 230026, China
}
\affiliation{
CAS Center for Excellence in Quantum Information and Quantum Physics, Hefei, 230026, China
}
\affiliation{
Hefei National Laboratory, University of Science and Technology of China, Hefei, 230088, China
}
\author{Sheng Liu}
\email{shengliu@ustc.edu.cn}
\affiliation{
 CAS Key Laboratory of Quantum Information,University of Science and Technology of China, Hefei, 230026, China
}
\affiliation{
 CAS Center for Excellence in Quantum Information and Quantum Physics, Hefei, 230026, China
}
\affiliation{
Hefei National Laboratory, University of Science and Technology of China, Hefei, 230088, China
}
\author{Yong-Sheng Zhang}
\email{yshzhang@ustc.edu.cn}
\affiliation{
 CAS Key Laboratory of Quantum Information,University of Science and Technology of China, Hefei, 230026, China
}
\affiliation{
 CAS Center for Excellence in Quantum Information and Quantum Physics, Hefei, 230026, China
}
\affiliation{
Hefei National Laboratory, University of Science and Technology of China, Hefei, 230088, China
}
\date{\today}

\begin{abstract}
We study the Faraday patterns generated by spin-orbit coupling induced parametric resonance in a spinor Bose-Einstein condensate with repulsive interaction. The collective elementary excitations of the Bose-Einstein condensate, including density waves and spin waves, are coupled as the result of the Raman-induced spin-orbit coupling and a quench of the relative phase of two Raman lasers without the modulation of any of the system's parameters. We observed several higher parametric resonance tongues at integer multiples of the driving frequency and investigated the interplay between Faraday instabilities and modulation instabilities when we quench the spin-orbit coupled Bose-Einstein condensate from zero-momentum phase to plane-wave phase. If the detuning is equal to zero, the wave number of combination resonance barely changes as the strength of spin-orbit coupling increases. If the detuning is not equal to zero after a quench, a single combination resonance tongue will split into two parts.
\end{abstract}

\maketitle

\section{\label{sec:level1}Introduction}
Spontaneous pattern formation is an important universal phenomenon in chemistry, biology, and physics \cite{Cross}. The pattern formation in a classical driving system was studied by Faraday in 1831 \cite{Faraday} using different kinds of liquid. He found that when the driving frequency exceeds the critical value, the surface of the liquid becomes spatially modulated, which is related to the parametric instability. Since the realization of Bose-Einstein condensate (BEC), one can study hydrodynamics of quantum fluid in experiment. This new type of nonlinear quantum fluid shows some different features from the classical counterpart. A BEC of the cold atoms manifests much more interesting nonlinear phenomena due to its quantum nature, such as solitons \cite{Muryshev, Kartashov}, vortices \cite{Raman, Neely, White, Weiler, Navon} and Faraday waves \cite{Staliunas, Engels, Zhang}. The tunable cold atom system provides a promising platform to investigate these fascinating phenomena. The nonlinear dynamics of BECs stem from the collision interaction between the atoms, which can be adjusted by exploiting the Feshbach resonances \cite{Theis}. The Faraday patterns of a BEC will emerge if one periodically drives the strength of nonlinearity \cite{Staliunas, Zhang, Nath, Abdullaev1, Abdullaev2, Okazaki}, or if the trap confinement of the BEC is periodically modulated \cite{Engels, Modugno, Balaz, Hernandez}, in which the nonlinear interaction is effectively oscillated. 

Realization of the synthetic spin-orbit (SO) coupled BEC \cite{Lin, Wu} enables study of some interesting phenomena in condensed matter systems, such as spin-Hall effect, supersolidity and topological insulators \cite{Zhai}. Raman-induced SO coupling can be generated by two counter propagating beams of laser light along the $x$ direction of an initially confined BEC in a typical experiment. By tuning the system's parameter, the ground state of the SO-coupled BEC will be in different phases, such as plane-wave phase, zero-momentum phase, or stripe phase. The spontaneous pattern formation of the SO-coupled BEC has been studied in Refs. \cite{Bhat, Li, Liu, Yi}, where the patterns are induced by modulation instabilities, a moving barrier, or the Kibble-Zurek mechanism, etc. However, investigation of spontaneous pattern formation induced by Faraday instabilities in the SO-coupled BEC is still lacking. 

In a recent paper \cite{Chen}, the authors found that the Faraday patterns can also be excited from a quench of the phase of the Rabi coupling without any modulation of the system's parameters. The effective modulation of interaction can be realized when the interaction coefficients satisfy $\textsl{g}_{12}^2\neq\textsl{g}_{11}\textsl{g}_{22}$ and the average populations of two hyperfine states of the spinor BEC experience Rabi oscillation. However, according to the simulation in Ref. \cite{Chen}, it takes a long time for the Faraday patterns to emerge and in a practical experiment, the tuning range of $|\textsl{g}_{12}^2-\textsl{g}_{11}\textsl{g}_{22}|$ is limited. Motivated by Ref. \cite{Chen}, here we show that the Faraday patterns in a SO-coupled BEC $(\textsl{g}_{ij}[i,j=1,2]=\textsl{g})$ can be induced by the Raman-induced SO coupling and a quench of relative phase of two Raman lasers without modulating any of the system's parameters such as the external potential, effective interaction, etc. In a typical experiment, the strength of SO coupling is highly tunable and can be adjusted by changing the angle between the two incident Raman lasers. One can also effectively tune the strength of SO coupling by periodic modulation of the power of the Raman lasers \cite{YPZhang, YPZhang2}. The spontaneous formation of the Faraday patterns is the direct consequence of the coupling of two types of collective elementary excitations (spin waves and density waves) due to the SO coupling and the Rabi oscillation induced by the quench. In our numerical simulation, we can observe the Faraday patterns emerge in a much shorter time. In the resonance diagram, we observed several higher resonance tongues at integer multiples of the driving frequency. If the detuning is not equal to zero after the quench, the resonance tongues will split into two parts as the strength of SO coupling increases. After quenching the SO-coupled BEC from zero-momentum phase to plane-wave phase, we can observe the coexistence of modulation instabilities and Faraday instabilities. 

Our paper is organized as follows. In Sec. \ref{sec:level2} we derive the coupled Gross-Pitaevskii (GP) equations of the SO-coupled BEC and its analytical solutions. In Sec. \ref{sec:level3} we derive the resonance conditions of our system. Section \ref{sec:level4} presents the instability analysis and the numerical results of integrating the GP equations. In Sec. \ref{sec:level5} we discuss more general cases. Section \ref{sec:level6} is our conclusion.

\section{\label{sec:level2}System model}

We consider a two-dimensional homogeneous BEC with Raman-induced SO coupling along the $x$ direction at zero temperature under the mean-field description. This two-dimensional geometry can be realized by applying a strong harmonic trapping potential of frequency $\omega_z$ in the $z$ direction. As we focus on the homogeneous case, the external potential in two dimensions is $V(x,y)=0$. The two incident counter propagating lasers act on the atoms at some angle with the $\textit{x}$ direction to synthesize the Rashba \cite{Bychkov} and the Dresselhaus \cite{Dresselhaus} SO-coupling interaction with equal contributions. Using the rotation approximation, the single particle Hamiltonian of interacting SO coupled BEC can be written as \cite{Lin} $H_{sp}=\hat{\bf p}^2/(2m)+\delta E\sigma_z/2+\hbar\Omega\cos(2k_r x-\delta\omega t+\phi)\sigma_x/2-\hbar\Omega\sin(2k_r x-\delta\omega t+\phi)\sigma_y/2$, where $\hat{\bf p}$ is the two-dimensional momentum operator, $m$ is the mass of the cold atoms, $\delta E$ is the energy level splitting, $\Omega$ is the strength of Raman coupling, $\delta\omega$ is the detuning of two Raman lasers, $k_r$ is the projected wavenumber of Raman lasers in the $x$ direction and $\phi$ is the relative phase of two Raman lasers. In the corotating frame of the effective field, under the unitary rotation $U = {\rm exp}[i(k_r x-\delta\omega t/2)\sigma_z]$, $H_{sp}^r = U^{\dagger}H_{sp}U-i\hbar U^{\dagger}\partial U/\partial t$ \cite{YPZhang}. The total mean field Hamiltonian becomes

\begin{subequations}\label{1}
\begin{gather}
H=H_{sp}^r+H_{int},\\
H_{sp}^r
=
  \begin{pmatrix}
     \dfrac{(\hat{\bf p}+\hbar k_r\hat{\bf e}_x)^2}{2m}+\dfrac{\hbar\Delta}{2} & \dfrac{\hbar\Omega}{2}e^{i\phi}\\
     \\
     \dfrac{\hbar\Omega}{2}e^{-i\phi} & \dfrac{(\hat{\bf p}-\hbar k_r\hat{\bf e}_x)^2}{2m}-\dfrac{\hbar\Delta}{2}
  \end{pmatrix},\\
H_{int}=
  \begin{pmatrix}
     \textsl{g}_{11}{\mid}{\varPsi_1}{\mid}^2+\textsl{g}_{12}{\mid}{\varPsi_2}{\mid}^2 & 0\\
     0 & \textsl{g}_{21}{\mid}{\varPsi_1}{\mid}^2+\textsl{g}_{22}{\mid}{\varPsi_2}{\mid}^2
  \end{pmatrix},
\end{gather}
\end{subequations}
where $\Delta=\delta E/\hbar-\delta\omega$ is the two-photon detuning, $\textsl{g}_{ij}(i,j=1,2)$ is the interaction coefficients, $\varPsi_1=\varPsi_1({\bf r})$ and $\varPsi_2=\varPsi_2({\bf r})$ represent the macroscopic condensate's wave functions where ${\bf r}=(x,y)$ is the two-dimensional spatial vector and $\Omega$ is the strength of Raman coupling. The strength of SO coupling is $\hbar k_r/m$. Diagonalizing the single particle Hamiltonian (1b), we can obtain the dispersion relationship 
\begin{equation}\label{2}
E_\pm=\frac{p^2}{2m}+\frac{\hbar^2 k_r^2}{2m}\pm\sqrt{(\frac{\hbar k_r p_x}{m}+\frac{\hbar\Delta}{2})^2+\frac{\hbar^2\Omega^2}{4}},
\end{equation}
\begin{figure}[t]
    \includegraphics[width=\textwidth/2,height=5cm]{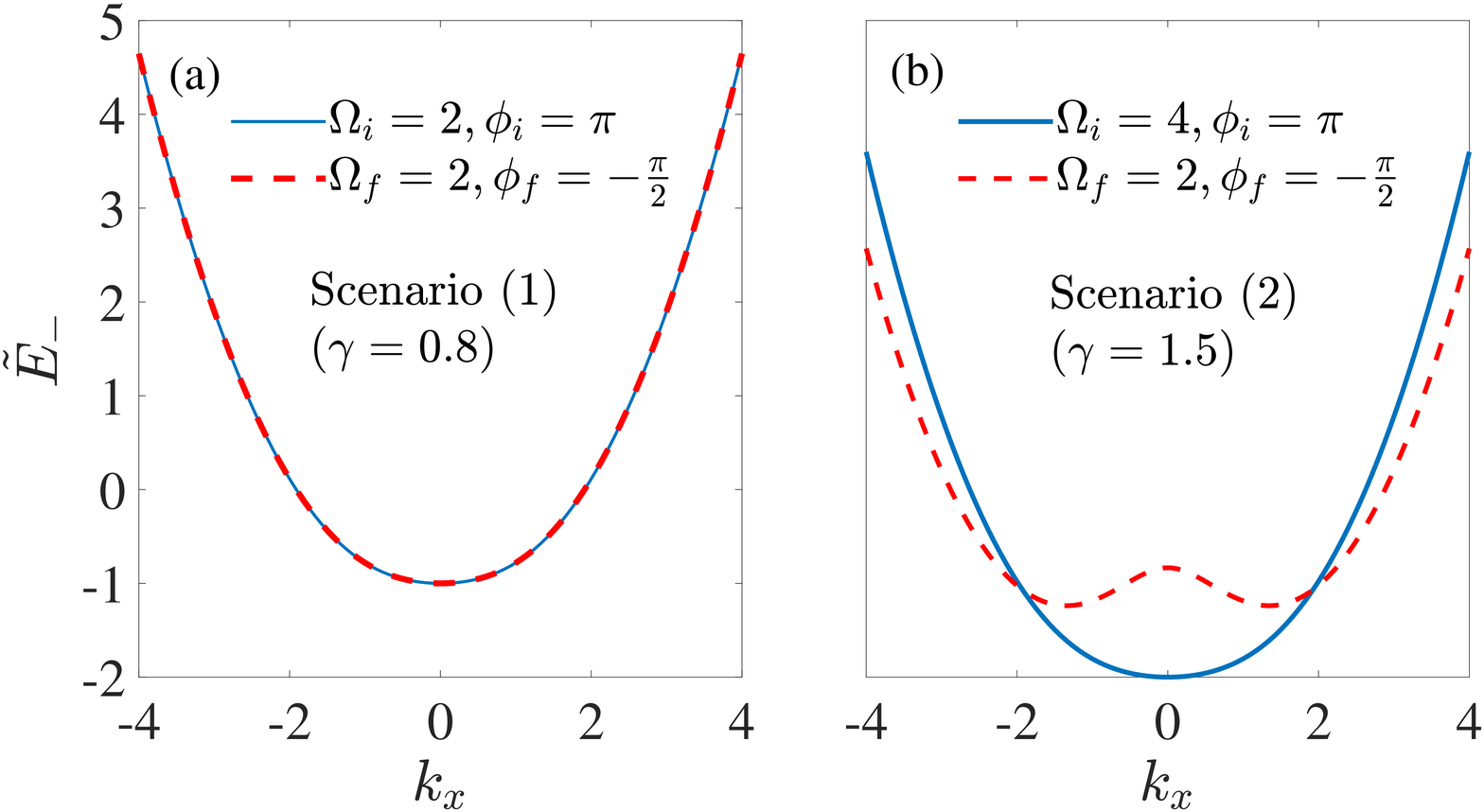}
    \caption{Dimensionless dispersion relation $\tilde{E}(\Delta=0)-=k_x^2/2-\sqrt{\gamma^2k_x^2+\Omega^2/4}$ with $\Omega=\Omega_i$ (solid line) and $\Omega=\Omega_f$ (dash line) corresponding to situations of before and after a quench, respetively. Circles in the figure mean the initial states before a quench located at $k_x=0$ before the quench, which indicates the SO-coupled BEC is in zero-momentum phase. (a) An example of both $\Omega_i$ and $\Omega_f$ are larger than $\Omega_c$. (b) An example of $\Omega_i>\Omega_c$ and $\Omega_f<\Omega_c$. }
    \label{fig_1}
\end{figure}
where $E_-$ is the ground state energy of the SO-coupled BEC. The corresponding eigenfunctions are
\begin{equation}\label{3}
\phi_+=e^{i{\bf p_+\cdot r}/{\hbar}}
\begin{pmatrix}
\cos(\theta/2)\\
\sin(\theta/2)
\end{pmatrix},
\phi_-=e^{i{\bf p_-\cdot r}/{\hbar}}
\begin{pmatrix}
\sin(\theta/2)\\
-\cos(\theta/2)
\end{pmatrix},
\end{equation}
where $\sin\theta=\hbar\Omega/\sqrt{4(\hbar p_x k_r/m-\hbar\Delta/2)^2+\hbar^2\Omega^2}$.

The time evolution of the spinor BEC is governed by the Schr\"{o}dinger equation
\begin{equation}\label{4}
i\hbar\frac{\partial}{\partial t}
\begin{pmatrix}
\varPsi_1({\bf r},t)\\
\varPsi_2({\bf r},t)
\end{pmatrix}
=H
\begin{pmatrix}
\varPsi_1({\bf r},t)\\
\varPsi_2({\bf r},t)
\end{pmatrix}.
\end{equation}
Substituting Eqs. (\ref{1}) into Eq. (\ref{4}), using the length unit $x_0=\sqrt{\hbar/m\omega_z}$, the time unit $t_0=1/\omega_z$, and the energy unit $e_0=\hbar\omega_z$, we can obtain the coupled dimensionless GP equations,
\begin{subequations}\label{5}
\begin{eqnarray}
i\frac{\partial \tilde{\varPsi}_1(\tilde{{\bf r}},\tilde{t})}{\partial \tilde{t}} & = & -\frac{\tilde{\nabla}^2 \tilde{\varPsi}_1(\tilde{{\bf r}},\tilde{t})}{2} - i\gamma\frac{\partial \tilde{\varPsi}_1(\tilde{{\bf r}},\tilde{t})}{\partial \tilde{x}} 
\nonumber \\      
& & + \tilde{\textsl{g}}_{11}{\mid} {\tilde{\varPsi}_1(\tilde{{\bf r}},\tilde{t})}{\mid}^2\tilde{\varPsi}_1(\tilde{{\bf r}},\tilde{t}) + \tilde{\textsl{g}}_{12}{\mid}{\tilde{\varPsi}_2(\tilde{{\bf r}},\tilde{t})}{\mid}^2\tilde{\varPsi}_1(\tilde{{\bf r}},\tilde{t})
\nonumber \\
& & +\frac{\tilde{\Delta}}{2}\tilde{\varPsi}_1(\tilde{{\bf r}},\tilde{t})+\frac{\tilde{\Omega}}{2}e^{i\phi}\tilde{\varPsi}_2(\tilde{{\bf r}},\tilde{t}),
\nonumber \\ \\
i\frac{\partial \tilde{\varPsi}_2(\tilde{{\bf r}},\tilde{t})}{\partial \tilde{t}} & = & -\frac{\tilde{\nabla}^2 \tilde{\varPsi}_2(\tilde{{\bf r}},\tilde{t})}{2} + i\gamma\frac{\partial \tilde{\varPsi}_2(\tilde{{\bf r}},\tilde{t})}{\partial \tilde{x}}
\nonumber \\
& & + \tilde{\textsl{g}}_{22}{\mid}{\tilde{\varPsi}_2(\tilde{{\bf r}},\tilde{t})}{\mid}^2\tilde{\varPsi}_2(\tilde{{\bf r}},\tilde{t}) + \tilde{\textsl{g}}_{21}{\mid}{\tilde{\varPsi}_1(\tilde{{\bf r}},\tilde{t})}{\mid}^2\tilde{\varPsi}_2(\tilde{{\bf r}},\tilde{t})
\nonumber \\
& & -\frac{\tilde{\Delta}}{2}\tilde{\varPsi}_2(\tilde{{\bf r}},\tilde{t})+\frac{\tilde{\Omega}}{2}e^{-i\phi}\tilde{\varPsi}_1(\tilde{{\bf r}},\tilde{t}).
\nonumber \\
\end{eqnarray}
\end{subequations}
Here $\tilde{\nabla}$ denotes dimensionless two-dimensional derivative and $\gamma=k_rx_0$ is the dimensionless strength of SO coupling; other variables and parameters with tilde are also dimensionless. We will omit all tildes in the equations \eqref{5} for convenience in our following discussions. After above treatment, the dimensionless dispersion relation $\tilde{E}_-(\Delta=0)=k_x^2/2-\sqrt{\gamma^2k_x^2+\Omega^2/4}$ is shown in Fig. \ref{fig_1}.

In the case of $\Delta = 0$, when $\Omega>\Omega_c=2\gamma^2$, the ground state of the SO-coupled BEC is in zero-momentum phase; while if $\Omega<2\gamma^2$, the ground state of the SO coupled BEC is in plane-wave phase, which corresponds to the dressed spin states $\ket{\downarrow^{\prime}}=\cos(\theta/2)\ket{\uparrow}-\sin(\theta/2))\ket{\downarrow}$ and $\ket{\uparrow^{\prime}}=\sin(\theta/2)\ket{\uparrow}-\cos(\theta/2)\ket{\downarrow}$, where $\sin\theta=\Omega/\sqrt{4\gamma^2 k_x^2+\Omega^2}$, $\ket{\uparrow}$ and $\ket{\downarrow}$ are the bare spin states of SO coupled BEC. 

We consider two kinds of quenching scenarios in the following: $(1)$ When $t=0$, we prepare the ground state of the SO coupled BEC in the zero-momentum phase ($\Omega_i>\Omega_c$) and the relative phase of two lasers $\phi=\pi$. When $t>0$, we suddenly quench the system by shifting the relative phase of two lasers to $\phi\neq\pi$ and keep the strength of Raman coupling unchanged $(\Omega_f=\Omega_i)$. $(2)$ We can quench both the relative phase of two lasers to $\phi\neq\pi$ and the strength of Raman coupling to $\Omega_f<\Omega_c$ simultaneously, after which the SO-coupled BEC will enter the plane-wave phase. After the quench, the initial state will not be in the ground state of the new Hamiltonian, it will start to evolve. Figure \ref{fig_1} shows two kinds of scenarios before and after the quench. In the present paper, we set $\textsl{g}_{ij}(i,j=1,2)=\textsl{g}=1.0$. 

In the following sections, we set the detuning $\Delta$ to zero and $\varPsi_1({\bf r},0)=\varPsi_2({\bf r},0)=$ const but quench the relative phase of two lasers from $\phi_i =\pi$ to $\phi_f=-\pi/2$ at $t>0$ for convenience in our discussions. We can obtain the following uniform solution from the coupled equations \eqref{5}

\begin{subequations}\label{6}
    \begin{gather}
       \varPsi_1(t) = e^{-i\textsl{g}t}\psi_1(t),\\
       \varPsi_2(t) = e^{-i\textsl{g}t}\psi_2(t),
    \end{gather}
\end{subequations}
where
\begin{subequations}\label{7}
    \begin{gather}
    \psi_1(t) = \frac{1}{\sqrt{2}}({\rm cos}(\Omega_f t/2)-{\rm sin}(\Omega_f t/2),\\
    \psi_2(t) = \frac{1}{\sqrt{2}}({\rm cos}(\Omega_f t/2)+{\rm sin}(\Omega_f t/2).
    \end{gather}
\end{subequations}
More general cases will be discussed in Sec. \ref{sec:level5}.
\section{\label{sec:level3} parametric resonance induced by Spin-orbit coupling}

In order to investigate the dynamics of excitations, we can add small fluctuations $\delta \varPsi_1$ and $\delta \varPsi_2$ to the uniform solution \eqref{6} and obtain
\begin{subequations}\label{8}
    \begin{gather}
    \varPsi_1^f(t) = e^{-i\textsl{g}t}(\psi_1(t)+\delta \varPsi_1),\\
    \varPsi_2^f(t) = e^{-i\textsl{g}t}(\psi_2(t)+\delta \varPsi_2).
    \end{gather}
\end{subequations}
Here we use the transformation in Ref. \cite{Chen} and define 
\begin{equation}\label{9}
\begin{pmatrix}
\delta\varPsi_d\\
\\
\delta\varPsi_s
\end{pmatrix}
=
\begin{pmatrix}
\psi_1(t) & \psi_2(t)\\
\\
-\psi_2(t) & \psi_1(t)  
\end{pmatrix}
\begin{pmatrix}
\delta\varPsi_1\\
\\
\delta\varPsi_2
\end{pmatrix},
\end{equation}
where $\delta\varPsi_d$ and $\delta\varPsi_s$ are the density and the spin fluctuations, respectively. Substituting Eqs. (\ref{8}) and (\ref{9}) into the coupled GP equations (\ref{5}) and neglecting the second-order and the third-order terms of $\delta\varPsi_d$ and $\delta\varPsi_s$, we obtain
\begin{subequations}\label{10}
\begin{eqnarray}
i\frac{\partial (\delta\varPsi_d)}{\partial t} & = & -\frac{\nabla^2}{2}\delta\varPsi_d + i\gamma \sin (\Omega_f t)\frac{\partial (\delta\varPsi_d)}{\partial x} 
\nonumber \\
& & +i\gamma \cos(\Omega_f t)\frac{\partial (\delta\varPsi_s)}{\partial x} + \textsl{g}(\delta\varPsi_d+\delta\varPsi^*_d),
\nonumber \\ \\
i\frac{\partial (\delta\varPsi_s)}{\partial t} & = & -\frac{\nabla^2}{2}\delta\varPsi_s
+ i\gamma \cos (\Omega_f t)\frac{\partial (\delta\varPsi_d)}{\partial x} 
\nonumber \\
& & - i\gamma \sin(\Omega_f t)\frac{\partial (\delta\varPsi_s)}{\partial x} .
\nonumber\\
\end{eqnarray}
\end{subequations}
It is now clear that the equations of fluctuations of density waves and spin waves become coupled with each other due to the SO coupling even if $\textsl{g}_{ij}(i,j=1,2)=\textsl{g}>0$ with driving frequency $\Omega_f$. If the strength of SO coupling is zero, Eqs. (\ref{10}) become decoupled, which is the same as the result in Ref. \cite{Chen}, where the Faraday pattern is absent although the relative phase of two Raman lasers is not equal to zero. In Ref. \cite{Chen}, the interatomic interaction effectively oscillates in time when the population of two states $\ket{\uparrow}$ and $\ket{\downarrow}$ oscillates since $\textsl{g}_{12}^2\neq\textsl{g}_{11}\textsl{g}_{22}$. In our case, if the system is in the zero-momentum phase, the interatomic interaction need not oscillate with time and the Faraday patterns will be introduced by SO coupling. 

For convenience, in the following discussion we set $k_y=0$ and $\gamma=0$. The Bogoliubov-de Gennes (BdG) Hamiltonian in momentum space becomes

\begin{equation}\label{11}
{\rm  \hat{H}_{BdG}}=
\begin{pmatrix}
k_x^2/2+\textsl{g} & \textsl{g} & 0 & 0\\
-\textsl{g} & -k_x^2/2-\textsl{g} & 0 & 0\\
0 & 0 & k_x^2/2 & 0\\
0 & 0 & 0 &  -k_x^2/2
\end{pmatrix}.
\end{equation}
The positive eigenvalues of the matrix are $\omega_d=\sqrt{\epsilon_k(\epsilon_k+2\textsl{g})}$, $\omega_s=\epsilon_k$, where $\epsilon_k=k_x^2/2$. Assume {\bf X} to be the fundamental solution of the differential equation $i\dot{{\bf x}}={\rm \hat{H}_{BdG}}{\bf x}$, where ${\bf x}=(\delta \varPsi_d, \delta\varPsi_d^*, \delta\varPsi_s, \delta\varPsi_s^*)^{\bf T}$. Then ${\bf X}={\rm {exp}}(-i{\rm \hat{H}_{BdG}}t)$. The monodromy matrix ${\bf F}_0={\bf X}({\rm T})={\rm {exp}}(-i{\rm H_{BdG}}{\rm T})$, where ${\rm T}=2\pi/\Omega_f$ is the period of driving. The multipliers $\rho_h$ are the eigenvalues of the monodromy matrix ${\bf F}_0$,
\begin{equation}\label{12}
\rho_h(h=d,s)={\rm exp}(\pm i\omega_h {\rm T})={\rm exp}(\pm i2\pi\omega_h/\Omega_f).
\end{equation}
The parametric resonance conditions are

\begin{subequations}\label{13}
\begin{gather}
2\omega_d= n\Omega_f, \quad n=1,2,3,...,\\
2\omega_s= n\Omega_f,\quad n=1,2,3,...,\\
\omega_d\pm\omega_s= n\Omega_f,\quad n=1,2,3,....
\end{gather}
\end{subequations}
The cases of (\ref{13}a) and (\ref{13}b) are called fundamental resonance while the cases of (\ref{13}c) are called combination resonance. The solutions of Eqs. \eqref{13} are 
\begin{subequations}\label{14}
\begin{gather}
k_d= \pm\sqrt{\sqrt{4g^2+n^2\Omega_f^2}-2}, \quad n=1,2,3,...\\
k_s= \pm\sqrt{n\Omega_f}, \quad n=1,2,3,...\\
k_{d+s}= \pm\frac{n\Omega_f}{\sqrt{n\Omega_f+\textsl{g}}}, \quad n=1,2,3,...
\end{gather}
\end{subequations}
Specifically, the case of $\omega_d-\omega_s=n\Omega_f$ has no real solution of $k_x$.  When $2\omega_h\neq n\Omega_f$ or $\omega_d\pm\omega_s\neq n\Omega_f$ ($n=1,2,...$), the multipliers are not double and are complex quantities lying on the complex unit circle, i.e., $|\rho_h|=1$. Therefore, the system is stable when $\gamma=0$. When $k_y=\Delta=0$ and for small $\gamma>0$, we can expand the monodromy matrix near $\gamma=0$ \citep{Seyranian} (see the Appendix). If at least one of the modules of the multipliers $|\rho_h|>1$, the system is dynamically unstable. If both modules of the multipliers $|\rho_{d,s}|<1$, the system will be asymptotically stable. The instability will occur in the vicinity of resonance points (\ref{14}a), (\ref{14}b), and (\ref{14}c), at which the multipliers are double. 
\section{\label{sec:level4} instability analysis and the numerical results ($\Delta=0$)}
We investigate the formations of Faraday patterns of the SO coupled BEC numerically by integrating the two dimensional GP Eqs. \eqref{5} using a pseudospectral method \cite{Dennis}. The spatial size of our system is 409.6 $\times$ 409.6. The wave function is discretized into a 2048 $\times$ 2048 mesh grid and the periodic boundary condition is adopted \cite{Chen}. The resolution in real space is $\Delta x=\Delta y=0.2$ and in momentum space it is $\Delta k =2\pi/409.6\approx0.015$ to ensure that the typical wave length and wave number of dynamics are larger than $\Delta x$ and $\Delta k$, respectively. The initial states are $\varPsi_1(t=0)=\varPsi_2(t=0)=1/\sqrt{2}$ plus small Gaussian noise. The resonance peaks can be recognized by the Fourier transformation of spatial wave functions. The resonance regions can also be calculated through Floquet analysis, by which we can calculate the unstable regions. Following the same procedure in Ref. \cite{Chen}, we assume the uniform solution
\begin{equation}\label{15}
\varPsi_j(t) = e^{-i\mu t}f_j(t) (j = 1,2),
\end{equation}
where $\mu$ is a constant and $f(t)$ is the complex periodic function with Rabi oscillation period ${\rm T}$. We add small excitations to the solution \eqref{15} and obtain
\begin{align}\label{16}
\varPsi_j({\bf r},t) = e^{-i\mu t}[f_j(t)+\delta\varPsi_j({\bf r},t)],\\
\delta\varPsi_j({\bf r},t)=u_j(t)e^{-i{\bf k}\cdot{\bf r}}+v_j^*(t)e^{i{\bf k}\cdot{\bf r}}.
\end{align}
Here ${\bf k}$ is the wave vector of the excitations. Substituting Eqs. \eqref{16} and (17) into GP equations \eqref{5}, we can obtain the coupled time dependent BdG equations
\begin{figure}[t]
    \includegraphics[width=10cm,height=6cm]{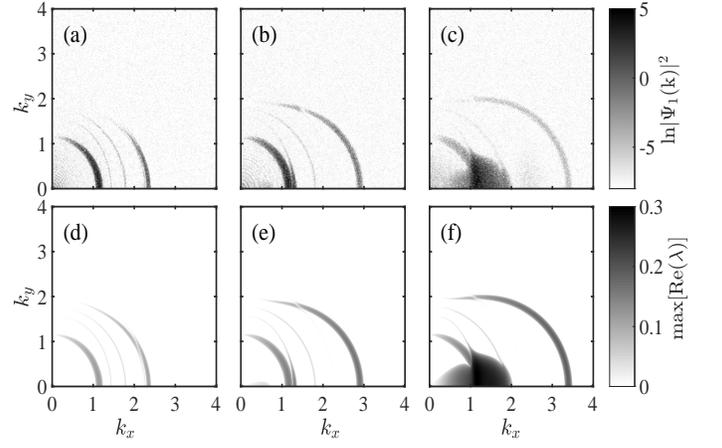}
    \caption{In the first column, $t = 52.5$ and $\gamma = 0.8$. In the second column, $t = 37.5$ and $\gamma = 1.2$. In the third column, $t = 22.5$ and $\gamma = 1.5$. (a)-(c) display the density distribution ${\rm log}(|\varPsi_1({\rm k})|^2)$ in the momentum space. (d)-(f) show the real part of Floquet exponent $\max[{\rm Re}(\lambda)]$ calculated by the time dependent BdG equations. The plots are symmetric about $k_x$ and $k_y$ axis, so we only show the results in the first quadrant. Here, $\Omega_f=2 $.}
    \label{fig_2}
\end{figure}
\begin{equation}\label{17}
i
\begin{pmatrix}
\dot{u_1}\\
\dot{v_1}\\
\dot{u_2}\\
\dot{v_2}
\end{pmatrix}
=
{\mathcal{L}}(t)
\begin{pmatrix}
u_1\\
v_1\\
u_2\\
v_2
\end{pmatrix}.
\end{equation}

Here 
\begin{subequations}\label{18}
\begin{equation}
{\mathcal{L}}(t)=
\begin{pmatrix}
 A_1+\gamma k_x & B_1 & C_1-i\dfrac{\Omega_f}{2} & C_2\\
-B_1^* & -A_1+\gamma k_x & -C_2^* & -C_1^*-i\dfrac{\Omega_f}{2}\\
C_1^*+i\dfrac{\Omega_f}{2} & C_2 & A_2-\gamma k_x & B_2\\
-C_2^* & -C_1+i\dfrac{\Omega_f}{2} & -B_2^* & -A_2-\gamma k_x
\end{pmatrix},
\end{equation}
\begin{gather}
A_1 = k^2_x/2+k^2_y/2-\mu+2\textsl{g}|f_1(t)|^2+\textsl{g}|f_2(t)|^2,\\
A_2 = k^2_x/2+k^2_y/2-\mu+2\textsl{g}|f_2(t)|^2+\textsl{g}|f_1(t)|^2,\\
B_1 = \textsl{g}f_1^2(t),\\
B_2 = \textsl{g}f_2^2(t),\\
C_1 = \textsl{g}f_1(t)f_2^*(t),\\
C_2 = \textsl{g}f_1(t)f_2(t).
\end{gather}
\end{subequations}
\begin{figure}[t]
    \includegraphics[width=9cm,height=5cm]{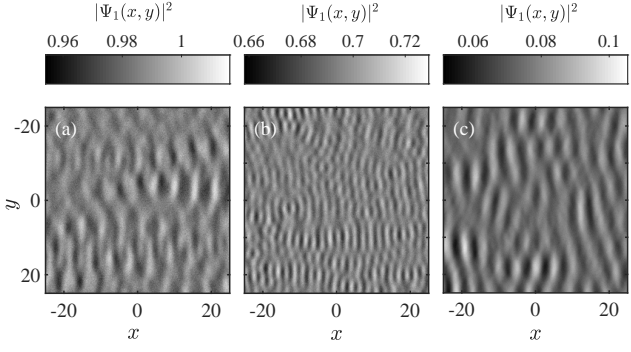}
    \caption{(a)-(c) are density distributions of $|\varPsi_1|^2$ in real space at $t = 52.5, 37.5$, and $22.5$ corresponding to $\gamma=0.8$, 1.2, and 1.5, respectively. Here $\Omega_f=2.0$.}
    \label{fig_3}
\end{figure}

The monodromy matrix ${\bf F}=\mathcal{T} {\rm {exp}}[-i\int_{0}^{\rm T}{\mathcal{L}}(t)dt]$, where $\mathcal{T}$ is the time-ordered operator. We calculate the monodromy matrix ${\bf F}$ numerically using the fourth order Runge-Kutta algorithm. The multipliers $\rho=e^{\lambda T}$ are obtained by diagonalizing the matrix ${\bf F}$. The real part of the Floquet exponent is defined as $\max[{\rm Re}(\lambda)]$. If $\max[{\rm Re}(\lambda)]> 0$, the system is unstable.

When $\Delta=0$ and $\Omega_f=2.0$ the results of time evolution of the GP equations are presented in Figs. \ref{fig_2}(a)-\ref{fig_2}(c), which display the density distribution of ${\rm log}(|\varPsi_1({\rm k})|^2)$ in the first quadrant of the momentum space, and in Fig. \ref{fig_3}, which displays the real space density distribution of $|\varPsi_1|^2$. The unstable regions identified by the value of $\max[{\rm Re}(\lambda)]$ are displayed in Fig. \ref{fig_2}(d)-\ref{fig_2}(f). We observe that the Faraday patterns appear earlier $(t<60t_0)$ than that in Ref. \cite{Chen} $(t>100t_0)$. In Figs. \ref{fig_3}(a)-\ref{fig_3}(c), the resulting spatial modulations of density in the $x$ direction are denser than that in the $y$ direction due to the larger wave number in the $k_x$ direction of the momentum space. We can also see that the resonance regions in the first row of Fig. \ref{fig_2} match very well with the unstable regions in the second row of Fig. \ref{fig_2}. We can clearly see four arcs in Figs. \ref{fig_2}(a) and \ref{fig_2}(b), where $\gamma=0.8$ and $\gamma=1.2$, respectively. The Raman-induced SO coupling is set to be in the $x$ direction so the resonance peaks are diminished in the $k_y$ direction.  We can also understand this from the BdG Hamiltonian \eqref{A1}, in which the driving terms are all multiplied by $\gamma$ and $k_x$. In the $k_y$ direction, $k_x=0$ and the driving terms become to zero, which indicates there is no coupling between the two types of excitations. Therefore, there is no resonance in this situation. The leftmost arcs in Figs. \ref{fig_2}(a) and \ref{fig_2}(b) correspond to the combination resonance conditions $\omega_d+\omega_s = \Omega_f$ and $k_{d+s}(n=1) = 2/\sqrt{3}\approx 1.15$. The other two fundamental resonance arcs ($n=1$) are too faint to be seen. This can be understood by calculating $\max[{\rm Re}(\lambda)]$. The results are shown in Figs. \ref{fig_2}(d)-\ref{fig_2}(e), in which the other two fundamental resonance arcs ($n=1$) are also extremely faint where $\max[{\rm Re}(\lambda)]\approx10^{-4}\sim10^{-2}$ at the resonance peaks. The rest three arcs correspond to the resonance conditions (\ref{13}a)-(\ref{13}c) ($n=2$). If we look at the rightmost arcs in Figs. \ref{fig_2}(d)-Fig. \ref{fig_2}(f) more closely, we can find another very thin and dim resonance arcs intercepting with the rightmost arcs. These thin arcs correspond to the higher combination parametric resonance, i.e., $\omega_d+\omega_s = 3\Omega_f$. In Figs. \ref{fig_2}(b) and \ref{fig_2}(c), the strength of SO coupling $\gamma>1$, and the system is in plane-wave phase. The modulation unstable region will start to grow and merge with the Faraday unstable region. In this case the SO-coupled BEC is in the dressed spin state $\ket{\uparrow^\prime}$ and $\ket{\downarrow^\prime}$. The effective interspecies $\textsl{g}_{{1^\prime}{2^\prime}}$ and intraspecies $\textsl{g}_{{1^\prime}{1^\prime}}$, $\textsl{g}_{{2^\prime}{2^\prime}}$ coefficients can be expressed in terms of bare interaction coefficient $\textsl{g}$ \cite{Li2}, i.e.,

\begin{figure}[t]
    \includegraphics[width=\columnwidth,height=4.7cm]{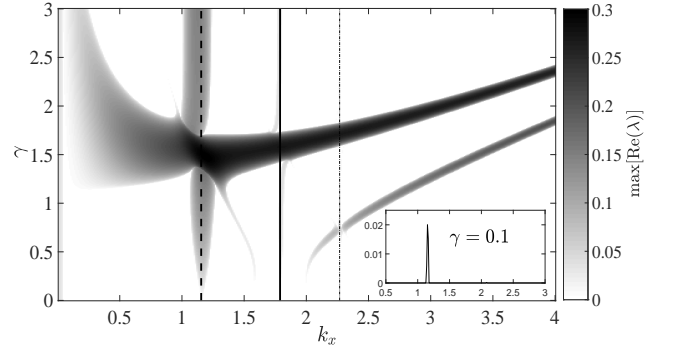}
    \caption{The diagram of the real part of Floquet exponent $\max[{\rm Re}(\lambda)]$ with $k_x$ and $\gamma$ when $\Delta=k_y=0$. The dashed line represents combination resonance ($n=1$) at $k_x =2/\sqrt{3}\approx 1.15$. The solid line represents combination resonance ($n=2$) at $k_x=4/\sqrt{5}\approx 1.79$. The dash dot line represents combination resonance ($n=3$) at $k_x=6/\sqrt{7}\approx 2.27$. Inset: The real part of Floquet exponent $\max[{\rm Re}(\lambda)]$ as a function of $k_x$ when $\gamma=0.1$. Here, $\Omega_f=2.0$.}
    \label{fig_4}
\end{figure}
\begin{figure}[t]
    \includegraphics[width=\columnwidth,height=5cm]{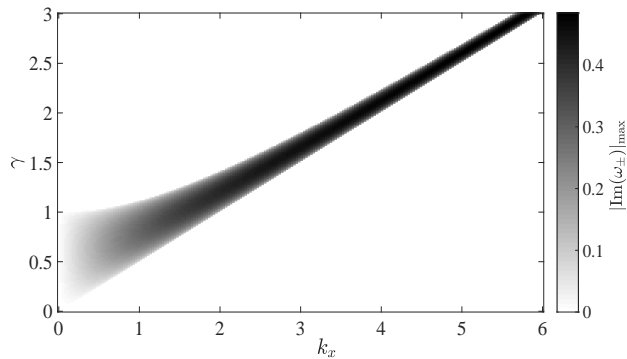}
    \caption{The diagram of $|{\rm Im}(\omega_\pm)|_{\max}$ about $k_x$ and $\gamma$ when $k_y=0$, $\textsl{g}=1$, and $\Omega_f\rightarrow0$.}
    \label{fig_5}
\end{figure} 

\begin{subequations}\label{19}
\begin{gather}
\textsl{g}_{{1^\prime}{1^\prime}}=\textsl{g},\\
\textsl{g}_{{2^\prime}{2^\prime}}=\textsl{g},\\
\textsl{g}_{{1^\prime}{2^\prime}}=2\textsl{g}-\cos^2\theta,
\end{gather}
\end{subequations}
where $\cos\theta=\gamma k_x/\sqrt{\gamma^2k_x^2+\Omega_f^2}$. Now the miscibility condition becomes 
\begin{equation}\label{20}
\eta= \frac{(\textsl{g}_{{1^\prime}{2^\prime}}^2-\textsl{g}_{{1^\prime}{1^\prime}}\textsl{g}_{{2^\prime}{2^\prime}})}{\textsl{g}_{{1^\prime}{1^\prime}}^2}=\frac{(2\textsl{g}-\cos^2\theta)^2-\textsl{g}^2}{\textsl{g}^2}.
\end{equation}
We set $\textsl{g}=1$ in the present paper so $\eta$ is always larger than zero, which indicates the dressed spin states are immiscible (modulation unstable) when the system enters into the plane-wave phase.  We also calculated the diagram of Floquet exponent ${\rm Re}(\lambda)_{\rm max}$ about $\gamma$ and $k_x$ ($k_y=0$). The result is shown in Fig. \ref{fig_4}. The dashed, solid, and dash-dot vertical lines represent three combination resonances ($n=1,2,3$), respectively. We can see the three combination resonances coincide with these vertical lines respectively, which indicates the combination resonance wave numbers $k_x$ barely change when the strength of SO coupling increases. The three resonance regions merge with the modulation unstable region when $\gamma$ is around $1.5$. The high resonance tongue ($n=3$) intercepts with the spin-wave unstable region $(n=2)$, other much higher resonance tongues ($n > 3$) are too dim to be seen. When the value of $\gamma$ is close to zero, the system is asymptotically stable except near the first combination resonance ($n=1$) (see the Appendix), so the rightmost four resonance peaks (including the $n=3$ resonance) in Fig. \ref{fig_4} disappear (see the inset in Fig. \ref{fig_4}). 

\begin{figure}[t]
    \includegraphics[width=\columnwidth,height=5cm]{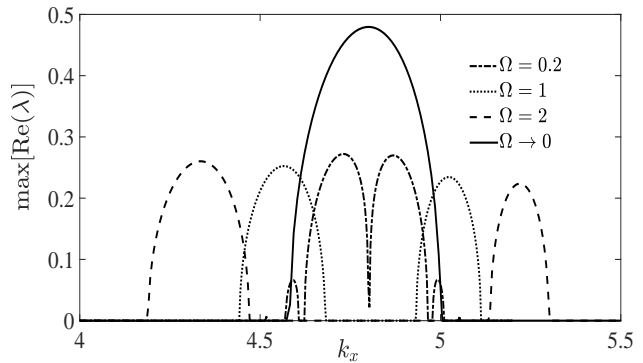}
    \caption{The value of $\max[{\rm Re}(\lambda)]$ as function of $k_x$ when $\gamma=2.5$ and $k_y=0$ with different values of $\Omega_f$. The solid line represents the analytical result of $|{\rm {Im}}(\omega_\pm)|_{\max}$ when $\Omega_f\rightarrow0$ .}
    \label{fig_6}
\end{figure}

In the limit of $\Omega_f\rightarrow 0$, the dispersion relation of excitations can be described by \cite{Bhuvaneswari}
\begin{equation}\label{21}
\omega_\pm^2=\frac{1}{2}(\Lambda_1\pm\sqrt{\Lambda_1^2+4\Lambda_2}),
\end{equation}
where $k^2 = k_x^2+k_y^2$ and
\begin{subequations}\label{22}
\begin{eqnarray}
\Lambda_1 & = & \frac{1}{2}k^2(k^2+2\textsl{g})+2k_x^2\gamma^2,
\nonumber \\ \\
\Lambda_2 & = & -k^2\Big[\gamma^2\Big(\gamma^2k_x^2 -\frac{1}{2}k^2(k^2+2\textsl{g})\Big)
\nonumber \\
& & +\frac{k^2}{16}(k^2+2\textsl{g})^2-\frac{k^2\textsl{g}^2}{4}\Big].
\nonumber \\
\end{eqnarray}
\end{subequations}
If the solutions of Eq. \eqref{22} are complex numbers, the amplitude of excitation grows exponentially with time. We can define $|{\rm {Im}}(\omega_\pm)|_{\max}$ as the growing rate. We calculate $|{\rm {Im}}(\omega_\pm)|_{\max}$ as a function of $k_x(k_y=0)$ and $\gamma$ numerically; the result is displayed in Fig. \ref{fig_5}. The location of modulation unstable regions moves in the positive $k_x$ direction as the strength of SO coupling $\gamma$ increases, similar to the behavior with the unstable region in Fig. \ref{fig_4} when $\gamma>1$. The rightmost region in Fig. \ref{fig_4} represents the spin wave excitation; it will merge with the modulation unstable region when the strength of Raman coupling $\Omega_f\rightarrow 0$, as we can see in Fig. \ref{fig_6}. 
\begin{figure}[t]
    \includegraphics[width=\columnwidth,height=5cm]{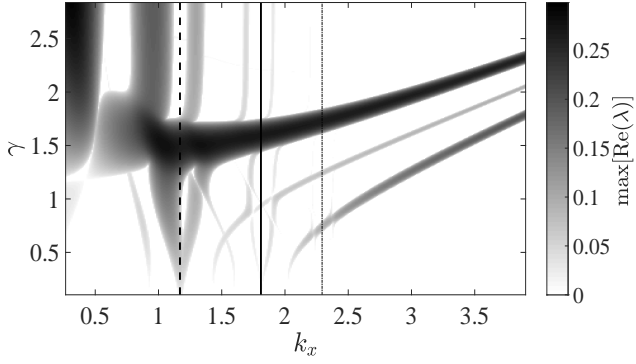}
    \caption{The diagram of real part of Floquet exponent $\max[{\rm Re}(\lambda)]$ with $k_x$ and $\gamma$ when $\Delta=0.4$ and $k_y=0$. The dashed,  solid and dash-dot lines represent the resonance condition $\omega_d+\omega_s=\sqrt{\Omega_f^2+\Delta^2}$, $\omega_d+\omega_s=2\sqrt{\Omega_f^2+\Delta^2}$ and $\omega_d+\omega_s=3\sqrt{\Omega_f^2+\Delta^2}$, respectively. Here, $\Omega_f=2.0$.}
    \label{fig_7}
\end{figure}
\section{\label{sec:level5} discussions of general cases}
\subsection{\label{subsec:level1}$\Delta \neq 0$}
When $\Delta\neq0$, the ground state of the SO coupled BEC will be in plane-wave phase. [In the present paper, we set $\textsl{g}_{ij}(i,j=1,2)=\textsl{g}=1$, so the stripe phase is absent in our system.]
We assume $\varPsi_1({\bf r},0)=\varPsi_2({\bf r},0)=$ const. We not only follow the quenching scenarios in Sec. II but also quench the detuning from $\Delta=0$ to $\Delta\neq 0$ simultaneously. Then we can obtain the following uniform solution of the coupled equations \eqref{5}

\begin{subequations}\label{23}
    \begin{gather}
       \varPsi_1(t) = e^{-i\textsl{g}t}\psi_1(t),\\
       \varPsi_2(t) = e^{-i\textsl{g}t}\psi_2(t),
    \end{gather}
\end{subequations}
where
\begin{subequations}\label{24}
    \begin{gather}
    \psi_1(t) = \frac{1}{\sqrt{2}}[{\rm cos}(\sqrt{\Omega_f^2+\Delta^2} t/2)-e^{i\xi}{\rm sin}(\sqrt{\Omega_f^2+\Delta^2} t/2)],\\
    \psi_2(t) = \frac{1}{\sqrt{2}}[{\rm cos}(\sqrt{\Omega_f^2+\Delta^2} t/2)+e^{i\xi}{\rm sin}(\sqrt{\Omega_f^2+\Delta^2} t/2)],\\
    \sin\xi = \frac{\Delta}{\sqrt{\Omega_f^2+\Delta^2}}.
    \end{gather}
\end{subequations}
Following the standard procedure in Sec. \ref{sec:level3}, we generalize the transformation in Eq. \eqref{9} and define 
\begin{equation}\label{25}
\begin{pmatrix}
\delta\varPsi_d\\
\\
\delta\varPsi_s
\end{pmatrix}
=
\begin{pmatrix}
\psi^*_1(t) & \psi^*_2(t)\\
\\
-\psi_2(t) & \psi_1(t)  
\end{pmatrix}
\begin{pmatrix}
\delta\varPsi_1\\
\\
\delta\varPsi_2
\end{pmatrix},
\end{equation}
where $\delta\varPsi_d$ and $\delta\varPsi_s$ are the density and the spin fluctuations, respectively. Substituting Eqs. (\ref{24}) and (\ref{25}) into the coupled GP equations \eqref{5} and neglecting the second-order and third-order terms of $\delta\varPsi_d$ and $\delta\varPsi_s$, we obtain
\begin{subequations}\label{26}
\begin{eqnarray}
i\frac{\partial (\delta\varPsi_d)}{\partial t} & = & -\frac{\nabla^2}{2}\delta\varPsi_d + i\gamma \cos{\xi} \sin (\sqrt{\Omega_f^2+\Delta^2} t)\frac{\partial (\delta\varPsi_d)}{\partial x} 
\nonumber \\
& & +i\gamma e^{-i\xi}\Big\{i\sin\xi +\cos \xi \cos (\sqrt{\Omega_f^2+\Delta^2} t)\Big\}
\nonumber \\
& & \times\frac{\partial (\delta\varPsi_s)}{\partial x} + \textsl{g}(\delta\varPsi_d+\delta\varPsi^*_d),
\nonumber \\ \\
i\frac{\partial (\delta\varPsi_s)}{\partial t} & = & -\frac{\nabla^2}{2}\delta\varPsi_s
+ i\gamma e^{-i\xi}\Big\{i\sin \phi
\nonumber \\
& & +\cos\xi \cos (\sqrt{\Omega_f^2+\Delta^2} t)\Big\}\frac{\partial (\delta\varPsi_d)}{\partial x} 
\nonumber \\
& & - i\gamma \cos \xi \sin(\sqrt{\Omega_f^2+\Delta^2} t)\frac{\partial (\delta\varPsi_s)}{\partial x} .
\nonumber\\
\end{eqnarray}
\end{subequations}
The resonance conditions near $\gamma=0$ are the same as Eqs. \eqref{13} if we replace $\Omega_f$ with $\sqrt{\Omega_f^2+\Delta^2}$. When the detuning $\Delta$ is not equal to zero, the system is in the plane wave phase. Its unstable behavior has some new features. We plot the diagram of the real part of Floquet exponent $\max[{\rm Re}(\lambda)]$ with $k_x$ and $\gamma$ in Fig. \ref{fig_7}. We observe that the combination resonance tongues will split into two parts as the strength of SO coupling $\gamma$ increases. The first ($n=1$) two fundamental resonance peaks reappear while they are absent when $\Delta=0$. 

\subsection{\label{subsec:level2}Arbitrary relative phase $\phi_f$}
\begin{figure}[t]
    \includegraphics[width=\columnwidth,height=4.7cm]{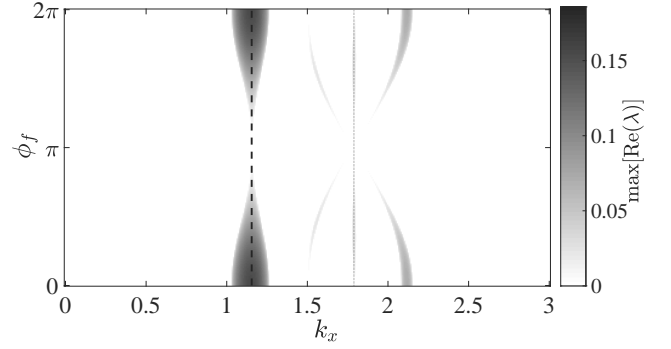}
    \caption{The diagram of real part of Floquet exponent $\max[{\rm Re}(\lambda)]$ with $k_x$ and $\phi$ when $\Delta=k_y=0$ and $\gamma=0.4$. The dashed line represents combination resonance ($n=1$) at $k_x =2/\sqrt{3}\approx 1.15$. The dotted line represents combination resonance ($n=2$) at $k_x=4/\sqrt{5}\approx 1.79$. Here, $\Omega_f=2.0$.}
    \label{fig_8}
\end{figure}
We assume $\Delta=0$, $\varPsi_1({\bf r},0)=\varPsi_2({\bf r},0)=$ const  and quench the relative phase from $\phi_i=\pi$ to an arbitrary $\phi_f\neq\pi$. Then we can obtain the uniform solution of the coupled equations \eqref{5}
\begin{subequations}\label{27}
    \begin{gather}
       \varPsi_1(t) = e^{-i\textsl{g}t}\psi_1(t),\\
       \varPsi_2(t) = e^{-i\textsl{g}t}\psi_2(t),
    \end{gather}
\end{subequations}
where
\begin{subequations}\label{28}
    \begin{gather}
    \psi_1(t) = \frac{1}{\sqrt{2}}[\cos(\frac{\Omega_f}{2}t)-ie^{i\phi_f}\sin(\frac{\Omega_f}{2}t)],\\
    \psi_2(t) = \frac{1}{\sqrt{2}}[\cos(\frac{\Omega_f}{2}t)-ie^{-i\phi_f}\sin(\frac{\Omega_f}{2}t)].
    \end{gather}
\end{subequations}
Using the generalized transformation \eqref{25}, following the same procedures in Sec. \ref{sec:level5}A, we can obtain the following coupled equations
\begin{subequations}\label{29}
\begin{eqnarray}
i\frac{\partial (\delta\varPsi_d)}{\partial t} & = & -\frac{\nabla^2}{2}\delta\varPsi_d - i\gamma \sin\phi_f\sin (\Omega_f t)\frac{\partial (\delta\varPsi_d)}{\partial x} 
\nonumber \\
& & + i\gamma[\cos(\Omega_f t)+i\cos\phi_f\sin(\Omega_f t)]\frac{\partial (\delta\varPsi_s)}{\partial x} 
\nonumber \\
& & + \textsl{g}(\delta\varPsi_d+\delta\varPsi^*_d),
\nonumber \\ \\
i\frac{\partial (\delta\varPsi_s)}{\partial t} & = & -\frac{\nabla^2}{2}\delta\varPsi_s
+ i\gamma[\cos(\Omega_f t)-i\cos\phi_f\sin(\Omega_f t)]
\nonumber \\
& & \times\frac{\partial (\delta\varPsi_d)}{\partial x} + i\gamma\sin\phi_f\sin(\Omega_f t)\frac{\partial (\delta\varPsi_s)}{\partial x}.
\nonumber\\
\end{eqnarray}
\end{subequations}
When $\gamma$ is close to zero, the corresponding resonance conditions are the same as Eqs. \eqref{13}. However, we should bear in mind that the initial state is the ground state of $H[\phi_f=(1+2m)\pi]$, $m=1,2,3,...$. If $\phi_f=(1+2m)\pi$, $m=1,2,3,...$, there will be no Faraday patterns. We plot the diagram of real part of the Floquet exponent of $\max[{\rm Re}(\lambda)]$ with $k_x$ and $\phi_f$ when $\Delta=k_y=0$ and $\gamma=0.4$ in Fig. \ref{fig_8}. The dashed line represents combination resonance ($n=1$) at $k_x =2/\sqrt{3}\approx 1.15$. The dotted line represents combination resonance ($n=2$) at $k_x=4/\sqrt{5}\approx 1.79$. Two fundamental resonances ($n=1$) and other higher resonance tongues ($n>2$) are too faint to be seen. We can see that the resonance regions are symmetric about $\phi=\pi$. When the relative phase $\phi=\pi$, there is no resonance region in the diagram, which means there is no pattern formation.
\section{\label{sec:level6} conclusion}
We have investigated the Faraday instability of homogeneous spinor BEC with SO coupling by a quench. We found that in the zero-momentum phase, the spatial patterns will emerge even if the interspecies and intraspecies interactions are the same. This is due to the fact that two fundamental excitations (excitations of spin waves and density waves) will be coupled with each other because of the SO coupling and the Rabi oscillations of the two components. We observe higher parametric resonance tongues at integer multiples of the driving frequency. The system is asymptotically stable except at the nearby of the first combination resonance. When we quench the SO coupled BEC from zero-momentum phase to plane-wave phase, the modulation instability begins to play the role of exponentially growing excitations. When the detuning is equal to zero, the wave number of the combination resonance barely changes as the strength of the SO coupling increases and for changes of relative phase of the two lasers. If the detuning is not equal to zero after a quench, a single combination resonance tongue will split into two parts as the strength of the SO coupling increases.  Recently, the BEC in an optical box potential was realized \cite{Gaunt}; we hope our theoretical calculation will inspire the observation of pattern formation of SO coupling BEC in the box potential. 

\begin{acknowledgements}
 This work has been supported by the National Natural Science Foundation of China (Grants No. 92065113) and the National Key R\&D Program. 
\end{acknowledgements}

\appendix* 

\section{stability of the system near $\gamma=0$}
\begin{figure}[t]
    \includegraphics[width=\columnwidth,height=5cm]{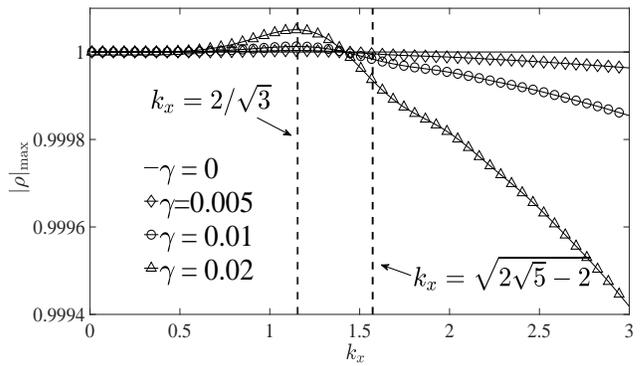}
    \caption{The maximum module of multipliers of the monodromy matrix ${\bf F}$ as the function of wave number $k_x$ with different strengths of SO coupling $\gamma$. Here, $\Omega_f=2.0$.}
    \label{fig_A1}
\end{figure}
When $k_y=\Delta=0$ and $\gamma\neq 0$, the Bogoliubov-de Gennes Hamiltonian ${\rm H}_{\rm BdG}(\gamma,t)$ in momentum space becomes
\begin{equation}\label{A1}
\begin{pmatrix}
\frac{k_x^2}{2}+{\rm S}+\textsl{g} & \textsl{g} & {\rm C} & 0\\
-\textsl{g} & -\frac{k_x^2}{2}+{\rm S}-\textsl{g} & 0 & {\rm C}\\
{\rm C} & 0 & \frac{k_x^2}{2}-{\rm S} & 0\\
0 & {\rm C} & 0 &  -\frac{k_x^2}{2}-{\rm S}
\end{pmatrix},
\end{equation}
where 
\begin{subequations}\label{A2}
\begin{gather}
{\rm S} = -\gamma k_x \sin(\Omega_f t),\\
{\rm C} = -\gamma k_x \cos(\Omega_f t).
\end{gather}
\end{subequations}
If $0<\gamma\ll1$, we can expand the monodromy matrix ${\bf F}=\mathcal{T} {\rm {exp}}[-i\int_{0}^{\rm T}{\rm H_{\rm BdG}}(\gamma,t)dt]$ near $\gamma=0$ and keep the terms of the first order of $\gamma$ \citep{Seyranian},
\begin{equation}\label{A3}
{\bf F}={\bf F}_0({\bf I}+{\bf A}\gamma),
\end{equation}
where ${\bf I}$ is the identity matrix and 
\begin{equation}\label{A4}
{\bf F}_0=
\begin{pmatrix}
{\rm P}({\rm T}) & {\rm Q}({\rm T}) & 0 & 0\\
{\rm Q}^*({\rm T}) & {\rm P}^*({\rm T}) & 0 & 0\\
0 & 0 & e^{-i\omega_s {\rm T}} & 0\\
0 & 0 & 0 & e^{i\omega_s {\rm T}}
\end{pmatrix},
\end{equation}
\begin{equation}\label{A5}
{\bf A}=\int_{0}^{\rm T}{\bf X}^{-1}\frac{\partial{\rm H_{BdG}(\gamma,t)}}{\partial\gamma}{\bf X}dt.
\end{equation}

After some calculation, we can reduce Eq. \eqref{A5} to a more simple form
\begin{equation}\label{A6}
{\bf A}=
\begin{pmatrix}
0 & 0 & M & N_1\\
0 & 0 & N_1^* & M^*\\
M^* & N_2^* & 0 & 0\\
N_2 & M & 0 & 0
\end{pmatrix},
\end{equation}
where
\begin{subequations}\label{A7}
\begin{gather}
M = -\int_{0}^{\rm T}k_x {\rm P}(t){\rm cos}(\Omega_f t)e^{-i\omega_s t}dt,\\
N_1 = -\int_{0}^{\rm T}k_x {\rm Q}(t){\rm cos}(\Omega_f t)e^{i\omega_s t}dt,\\
N_2 = -\int_{0}^{\rm T}k_x {\rm Q}(t){\rm cos}(\Omega_f t)e^{-i\omega_s t}dt,\\
{\rm P}(t) = {\rm cos}(\omega_d t)+i{\rm sin}(\omega_d t)(\epsilon_k+\textsl{g})/\omega_d,\\
{\rm Q}(t) = i\textsl{g}{\rm sin}(\omega_d t)/\omega_d.
\end{gather}
\end{subequations}

The multipliers $\rho$ are the eigenvalues of the matrix ${\bf F}$. We numerically calculate the maximum module of eigenvalues $|\rho|_{\max}$ with $\textsl{g}=\Omega_f/2=1.0$ and different values of $\gamma$. The result is presented in Fig. \ref{fig_A1}. $k_x = 2/\sqrt{3}$ corresponds to the wave number of combination resonance ($n=1$). $k_x = \sqrt{2\sqrt{5}-2}$ corresponds to the density waves fundamental resonance ($n=2$) . In Eqs. \eqref{14}, $k_d<k_{d+s}<k_s$ if $\textsl{g}=\Omega_f=1.0$. Thus when $k_x> \sqrt{2\sqrt{5}-2}$, the multipliers are always smaller than $1$ and the system is asymptotically stable.

\end{document}